\documentclass[aps,prb,reprint,superscriptaddress,amsmath,amssymb]{revtex4-2}

\usepackage{amsfonts,amsbsy,amsthm}
\usepackage{xcolor}
\usepackage{bm}
\usepackage{graphicx}
\usepackage{yhmath}
\usepackage{slashed}
\usepackage{mathtools}
\usepackage{ulem}
\usepackage[colorlinks=true,
            citecolor=black,
            linkcolor=black,
            urlcolor=black]{hyperref}

\def\XXint#1#2#3{{\setbox0=\hbox{$#1{#2#3}{\int}$ }
\vcenter{\hbox{$#2#3$ }}\kern-.6\wd0}}

\newcommand{\B}{{\bf B}}
\newcommand{\E}{{\bf E}}

\newcommand{\J}{{\bf J}}

\newcommand{\m}{{\bf m}}
\newcommand{\M}{{\bf M}}
\newcommand{\e}{{\bf e}}

\newcommand{\p}{{\bf p}}
\renewcommand{\j}{{\bf j}}

\renewcommand{\k}{{\bm{k}}}

\newcommand{\bsigma}{\boldsymbol{\sigma}}

\definecolor{bvio}{rgb}{0.54, 0.17, 0.89}

\definecolor{fgr}{rgb}{0.13, 0.55, 0.13}

\begin{document}

\title{In-plane magnetic response and Maki parameter of alternating-twist multilayers}

\author{Igor Vasilevskiy}
\email{igor.vasilevskiy@csic.es}

\affiliation{Quantum Advanced Research Center (QuARC), CSIC, E-28049 Madrid, Spain}
\affiliation{Instituto de Ciencia de Materiales de Madrid (ICMM), CSIC, E-28049 Madrid, Spain}

\author{Miguel S\'anchez S\'anchez}
\affiliation{Instituto de Ciencia de Materiales de Madrid (ICMM), CSIC, E-28049 Madrid, Spain}

\author{Khadija Challaouy}

\affiliation{Universidad Internacional Men\'endez Pelayo, E-28040 Madrid, Spain}

\author{Dionisios Margetis}

\affiliation{Department of Mathematics, and Institute for Physical Science
and Technology, University of Maryland, College Park, Maryland 20742, USA}

\author{Guillermo G\'omez-Santos}
\affiliation{Departamento de F\'{\i}sica de la Materia Condensada,
Instituto Nicol\'as Cabrera and Condensed Matter Physics Center (IFIMAC),
Universidad Aut\'onoma de Madrid, E-28049 Madrid, Spain}

\author{Tobias Stauber}
\email{tobias.stauber@csic.es}
\affiliation{Quantum Advanced Research Center (QuARC), CSIC, E-28049 Madrid, Spain}
\affiliation{Instituto de Ciencia de Materiales de Madrid (ICMM), CSIC, E-28049 Madrid, Spain}

\date{\today}

\begin{abstract}
We analytically study the orbital response of alternating-twist multilayer graphene to an in-plane magnetic field using the unitary transformation introduced by Khalaf \textit{et al.} [Phys.\ Rev.\ B \textbf{100}, 085109 (2019)]. This transformation maps an alternating-twist $N$-layer system onto $N/2$ decoupled twisted bilayer graphene (TBG) systems with distinct effective twist angles, together with a single decoupled layer for odd $N$, thereby generating a hierarchy of effective magic angles. For systems with an odd number of layers, we find that the orbital in-plane magnetic response is negligibly small. For even systems, we express the in-plane orbital susceptibility in terms of the corresponding TBG responses in the flat-band regime, which are large compared to the spin susceptibility and even diverge in the clean limit at charge neutrality near the magic angle. For the finite even-layer systems considered in this work, the in-plane magnetic response strongly depends on the effective magic angle within the hierarchy: the larger the twist angle, the smaller the total response. Moreover, we find a general relation between the outermost interlayer and total susceptibilities of the system when the corresponding effective TBG subsystem is in the flat-band regime. We finally introduce the in-plane Maki parameter as the ratio of the difference in orbital susceptibility between the normal and superconducting states to the paramagnetic Pauli susceptibility. For TBG, we find values up to 2 near the magic angle. Our analysis shows that for certain magic angles, the interpretation of Pauli-limit violation in alternating-twist multilayers requires taking into account the orbital contribution to the in-plane magnetic response.
\end{abstract}

\maketitle

\section{Introduction}
\label{sec:Intro}
The discovery of superconductivity in twisted bilayer graphene \cite{Cao_2018unconv} at the magic angle (MATBG) has attracted much attention by showing that flat-band engineering can induce unexpected phase transitions \cite{suarezmorell,bistritzer,Carr2017,Cao_2018,Chittari2018,Isobe2018,Kennes2018,Koshino2018,Zou2018,Gonzalez2019,Kang2019,Lian2019,Polshyn2019,Roy2019,Zhang2019,Cao2020,Chichinadze2020,Nuckolls2020,Saito2020,Lian2021,Saito2021,Xie2021,Jaoui2022,Dong2023,Chou2024,Rai2024,Wang2024,Zhou2024,Park2026}. The MATBG phase diagram shows notable parallels to what is observed in high-$T_{\rm{c}}$ superconductors, with the superconducting dome emerging in close proximity to an insulating phase \cite{Andrei20}. Moreover, the notably large ratio between the critical and Fermi temperatures places MATBG within the strong-coupling regime of known superconductors \cite{Yankowitz2019,Balents2020}. 

Superconductivity has since been reported in related graphene moir\'e systems, which include both commensurate \cite{Park21,Zeyu21,Cao21,Zhang2022,Kim23} and incommensurate structures \cite{Uri23,Xia2025}. In addition, several theoretical and experimental works have recently explored the electronic structure of incommensurate moiré and moiré-of-moiré systems, revealing that slowly varying supermoiré patterns and quasiperiodicity provide another route to correlated flat-band physics \cite{Mao2023,Guerci2024,Nakatsuji23,Hao2024}. In alternating-twist graphene multilayer systems, superconductivity is likewise expected because the Hamiltonian can be mapped onto decoupled twisted bilayer graphene (TBG) systems for an even number of layers, and onto decoupled TBG systems plus an additional decoupled single layer graphene (SLG) for an odd number of layers \cite{Khalaf19}. This mapping allows one to predict the magic angles for an arbitrary number of layers through $\theta_{k,m}^{N}=\beta_k^{N}\theta_m$, where $\beta_k^{N}=2\cos[\pi k/(N+1)],\; k=1,\dots,\lfloor N/2\rfloor$ \cite{Khalaf19}, and $\theta_m$ is the TBG magic angle. For example, this yields $\theta^3_{m}=\sqrt{2}\theta_m$, $\theta_{k,m}^4\in\{\varphi\theta_m, \varphi^{-1}\theta_m\}$, and $\theta_{k,m}^5\in \{\sqrt{3}\theta_m, \theta_m\}$ for $N=3,4,5$, respectively, where $\varphi=(1+\sqrt{5})/2$. In these systems, electrostatic effects and layer-dependent charge redistribution can play an important role \cite{Kolar2023}, influencing the effective band filling and response properties.

Although the systems exhibit certain similarities, the superconducting pairing mechanism may differ. One way to analyze this is by measuring the violation of the Pauli limit. This limit, derived from BCS theory, predicts the critical magnetic field that is needed to break superconductivity by aligning the spins of the two electrons that form the singlet Cooper pair. This Clogston-Chandrasekhar or Pauli limit is given by $B_{\rm{P}}=1.86\,T_{\rm{c}}$ (in tesla for $T_{\rm{c}}$ in kelvin) \cite{Clogston62,Chandrasekhar62}. 

In typical experiments, the magnetic field is applied in the in-plane direction to avoid additional orbital effects, which vanish in a purely two-dimensional structure. In fact, a violation of the Pauli limit by a factor of 2-3 was found in alternating-twist multilayers with $N=3,4,5$ \cite{Park22,Ledwith22}, 
corroborating the view that superconductivity is unconventional in these systems \cite{Lake2021,Christos22}. This interpretation was ultimately confirmed for the trilayer system through combined tunneling spectroscopy and transport measurements \cite{Park2026}.

The in-plane orbital susceptibility of TBG is intrinsically large \cite{Stauber18,Guerci21,Stauber23}, giving rise to exceptionally strong orbital magnetization responses \cite{He2020}. This mechanism is
distinct from the enhancement of the out-of-plane orbital susceptibility near Van Hove singularities recently observed in graphene on aligned
hBN moiré superlattices \cite{VallejoBustamante2023}, since the in-plane response is governed primarily by the counterflow currents associated with the
flat-band regime. As a consequence, the orbital contribution dominates over the spin susceptibility of Cooper pairs, so that no direct conclusion on the pairing symmetry can be drawn. In view of the mapping of alternating-twist multilayers onto effective TBG systems, it is therefore somewhat surprising that their in-plane orbital response can be significantly reduced and does not mask the spin susceptibility, particularly in the case with $N=4$.

In this paper, we analytically study the orbital response of alternating-twist graphene multilayers due to an in-plane magnetic field. Our approach is based on the unitary transformation introduced in~\cite{Khalaf19}, which allows us to express the response of a general $N$-layer system in terms of the corresponding effective TBG subsystems. We explicitly derive the response for the tetralayer ($N=4$) and pentalayer ($N=5$) systems, which illustrate the distinct behavior of even- and odd-layer structures. This extends substantially our previous results on the optical response for the particular case of the alternating-twist trilayer ($N=3$) \cite{Margetis24}. We further introduce and calculate the in-plane Maki parameter, which quantifies the relative importance of orbital and spin contributions to the magnetic response in the superconducting phase.

For trilayers, the small in-plane orbital magnetic response can be attributed to the mirror symmetry of the system \cite{Margetis24}. More generally, we find that alternating-twist multilayers with an odd number of layers exhibit a negligibly small orbital response, as explicitly demonstrated in this work for the pentalayer case. By contrast, multilayers with an even number of layers are not mirror symmetric and can display a large in-plane orbital magnetic response near the magic angle regime. However, the magnitude of this response strongly depends on the particular magic angle within the hierarchy $\theta_{k,m}^{N}$. For the tetralayer system, we show that the response at the larger magic angle, $\varphi\theta_m$, is strongly suppressed, allowing one to access the spin susceptibility of Cooper pairs without a dominant orbital contribution. Conversely, at the smaller magic angle, $\varphi^{-1}\theta_m$, the system exhibits a large orbital magnetic response comparable to that of TBG in the flat-band regime. Thus, different effective magic angles within the same alternating-twist multilayer can exhibit qualitatively distinct magnetic behavior, leading to different orbital corrections to the measured superconducting critical fields. Moreover, for even-layer alternating-twist systems, we propose a general relation between the total susceptibility and the outermost interlayer susceptibility in the flat-band regime, which we verify for the finite multilayer systems considered in this work. These results, together with the introduction of the in-plane Maki parameter, constitute the main highlights of the present paper.

The remainder of the paper is organized as follows. In Sec.~\ref{sec:micro}, we discuss our approach and introduce the layer-resolved conductivity tensor. We also define the magnetic field and dipole density in terms of layer-contrasted electric fields and currents, respectively. Section~\ref{sec:magnetic_response} is devoted to the calculation of the in-plane magnetic response. There, we express the response of the multilayer systems in terms of the corresponding effective TBG subsystems obtained through the unitary transformation and analyze the resulting magic-angle dependence. In Sec.~\ref{sec:Maki}, we address the superconducting phase and discuss the Pauli limit and the in-plane Maki parameter. Section~\ref{sec:conclusion} summarizes the main findings. The three appendixes provide additional details and results.

\section{Response theory for multilayers}
\label{sec:micro}
The electromagnetic response of layered two-dimensional systems to in-plane electric or magnetic fields can be decomposed into the sheet current responses of the individual layers, extending the approach of Ref.~\onlinecite{Margetis24}. These sheet currents are obtained from the layer-resolved Ohm's law, with dynamical conductivities obtained within linear response. We will argue that the magnitude of the equilibrium response
can then be inferred from the static ordered limit $\lim_{\omega\to0}\lim_{\mathbf{q}\to0}$ of the dynamical conductivities at charge neutrality. This argument relies on the observation that the equilibrium response, which requires the reverse order of limits, $\lim_{\mathbf{q}\to0}\lim_{\omega\to0}$, can be related to the present dynamical calculation, through a ``contact term" that only includes contributions from the Fermi surface, see Ref. \onlinecite{Stauber18b}. At charge neutrality, this contact term vanishes, and the two types of limits therefore coincide. Furthermore, we will argue that for the orbital magnetic susceptibility, the value at charge neutrality becomes representative of the entire band even at finite doping. 

Throughout this paper, we argue that the electromagnetic response of the multilayer can be expressed in terms of the response of TBG. For this purpose, we rely on the unitary transformation introduced by Khalaf \textit{et al.} \cite{Khalaf19}, which maps an alternating-twist multilayer with an even number of layers $N$ onto $N/2$ decoupled twisted bilayers. For TBG, it has been shown that near charge neutrality the magnetic response is approximately constant. Consequently, the magnetic response evaluated at $\mu=0$ in the present treatment should provide a reliable estimate of the equilibrium response of the multilayer systems considered here. Moreover, it sets the scale for the Fermi surface contribution to the orbital susceptibility, which is always paramagnetic.

We are particularly interested in the in-plane response around the flat-band regime characterized by the magic angles $\theta_{k,m}^N$. In TBG, the magnetic response in the clean limit at $\mu=0$ is paramagnetic and diverges algebraically as a function of the twist angle with $(\theta-\theta_m)^{-0.2}$ for $\theta>\theta_m$ \cite{Stauber23}. The divergence at the magic angle will be regularized in realistic systems, and we denote the resulting (finite) susceptibility by $\chi_{\mathrm{TBG}}$, which will serve as a reference scale for the equilibrium response of the multilayer systems.

\subsection{Layer-resolved Ohm's law and Kubo formula}
We consider general alternating-twist moir\'e multilayers, where the twist angle of layer $\ell$ is given by $\theta(\ell)=(-1)^{\ell}\theta/2$, with $0<\theta<\pi/2$ and $\ell=1,\dots,N$, and the interlayer distance is $a=3.4$\,\text{\AA}. While the formalism applies to arbitrary $N$, we present explicit analytical derivations for the tetralayer ($N=4$) and pentalayer ($N=5$) systems. For completeness, we also review the trilayer ($N=3$) case, previously discussed in Ref.~\onlinecite{Margetis24}, and provide susceptibility results for the hexalayer ($N=6$) and octalayer ($N=8$) systems in Appendix~C.

Ohm's law for these general layer-resolved systems in the frequency domain is given by~\cite{Stauber18,Stauber18b}
\begin{align}\label{eq:J-response}
\J_\ell=\sum_{\ell'=1}^N\bsigma^{\ell\ell'}\E_{\ell'},\quad \ell=1,\,\ldots,\,N\;,
\end{align}
where $\J_\ell$ and $\E_\ell$ denote the macroscopic surface current density and electric field in layer $\ell$, respectively. The $2\times2$ matrices $\bsigma^{\ell\ell'}(\omega)$ have elements defined by
\begin{align}\label{eq:sigma_ll'_nunu'}
\sigma^{\ell\ell'}_{\nu\nu'}(\omega)=i\frac{e^2}{\omega+i\delta}\chi^{\ell\ell'}_{\nu\nu'}(\omega+i\delta),
\end{align}
with $\delta\downarrow 0$ ensuring a retarded response. The current-current response function reads
\begin{align}\label{eq:resp-func}
\chi^{\ell\ell'}_{\nu\nu'}(\omega)=-\frac{i}{\hbar}\int_0^\infty dt\ e^{i\omega t}\langle[ j^\ell_\nu(t),j^{\ell'}_{\nu'}(0)]\rangle\;,
\end{align}
where $j^\ell_\nu(t)$ is the $\nu$-directed current operator ($\nu=x,y$) at layer $\ell$ in the interaction picture, and $\langle \cdot \rangle$ denotes the equilibrium average. In the following, we will also sometimes use the shortcut notation  $\langle\langle j^\ell_\nu(t),j^{\ell'}_{\nu'}\rangle\rangle=\chi^{\ell\ell'}_{\nu\nu'}$.

The alternating-twist geometry imposes certain symmetries on the total conductivity. The $2\times2$ matrices connecting layers with the same twist angle are proportional to the identity matrix:
\begin{align}
\bsigma^{\ell(\ell+2n)}=\sigma^{\ell(\ell+2n)}_0{\bf 1}\;,\quad \ell+2n\leq N,
\end{align}
where $\sigma^{\ell\ell'}_0$ is the longitudinal conductivity. Chirality is encoded in the off-diagonal entries of the conductivity tensor coupling the layers with opposite twist angles:
\begin{align}
\boldsymbol{\sigma}{}_{\pm}^{\ell(\ell+2n+1)}=
\sigma_{0}^{\ell(\ell+2n+1)}{\bf 1}
\pm i\sigma_{xy}^{\ell(\ell+2n+1)}\boldsymbol{\tau}_y,\notag\\
\ell + 2n + 1 \leq N,
\end{align}
where \(\boldsymbol{\tau}_y\) denotes the $y$-Pauli matrix. Moreover, time-reversal symmetry implies $\sigma^{\ell\ell'}_{\nu\nu'}=\sigma^{\ell'\ell}_{\nu'\nu}$. 

The total conductivity matrix for the trilayer system then reads \cite{Margetis24}
\begin{align}\label{eq:Sigma3}
\sigma_{\mathrm{tot}}&=
\begin{pmatrix}
\bsigma^{11} & \bsigma^{12}_+& \bsigma^{13} \\
\bsigma^{12}_- & \bsigma^{22}& \bsigma^{12}_-\\ 
\bsigma^{13} & \bsigma^{12}_+& \bsigma^{11}
\end{pmatrix}\;.
\end{align}
For $N=4$, this matrix becomes
\begin{align}\label{eq:Sigma4}
\sigma_{\mathrm{tot}}&=
\begin{pmatrix}
\bsigma^{11} & \bsigma^{12}_+& \bsigma^{13}&\bsigma^{14}_+ \\
\bsigma^{12}_- & \bsigma^{22}& \bsigma^{23}_-&\bsigma^{13} \\ 
\bsigma^{13} & \bsigma^{23}_+& \bsigma^{22}&\bsigma^{12}_+ \\ 
\bsigma^{14}_- & \bsigma^{13}& \bsigma^{12}_-&\bsigma^{11}  
\end{pmatrix}\;.
\end{align}
For $N=5$, we have 
\begin{align}\label{eq:Sigma5}
\sigma_{\mathrm{tot}}&=
\begin{pmatrix}
\bsigma^{11} & \bsigma^{12}_+& \bsigma^{13}&\bsigma^{14}_+&\bsigma^{15} \\
\bsigma^{12}_- & \bsigma^{22}& \bsigma^{23}_-&\bsigma^{24}&\bsigma^{14}_- \\ 
\bsigma^{13} & \bsigma^{23}_+& \bsigma^{33}&\bsigma^{23}_+&\bsigma^{13} \\ 
\bsigma^{14}_- & \bsigma^{24}& \bsigma^{23}_-&\bsigma^{22}&\bsigma^{12}_- \\ 
\bsigma^{15} & \bsigma^{14}_+& \bsigma^{13}&\bsigma^{12}_+&\bsigma^{11} 
\end{pmatrix}\;.
\end{align}
One can see that the alternating-twist geometry and time-reversal symmetry constrain the layer-resolved conductivity, reducing it to a limited number of independent response functions, i.e., 4, 6, and 9 for $N=3,4$, and 5, respectively. The conductivity tensor for higher values of $N$ can be 
constructed analogously by applying the same symmetry arguments.

\subsection{Electric and magnetic fields}
\label{sec:MagneticMoment}
Until now, we have only considered Ohm's law involving layer-resolved sheet current densities and electric fields. We can now define the average electric field
\begin{align}
\E^\parallel&=\frac{1}{N}\sum_{\ell=1}^N\E_\ell\;.
\end{align}
In order to define the layer differences, we will introduce magnetic quantities by discretizing the two equations $-\partial_t\B=\boldsymbol{\nabla}\times\E\,\to\,\partial_z\e_z\times\E$ and $\j=\boldsymbol{\nabla}\times\m\,\to\,\partial_z\e_z\times\m$. From the discrete (layer-resolved) version of the Maxwell-Faraday law we get the following relations for the average magnetic field between layers $\ell$ and $\ell'$:
\begin{align}
i\omega a(\ell-\ell')\B^\parallel_{\ell\ell'}&=\e_z\times(\boldsymbol \E_{\ell}-\boldsymbol\E_{\ell'})\;.\label{eq:B-par-n}
\end{align}
Note that a constant in-plane magnetic field within the sample is given by $\B^\parallel=\B^\parallel_{(\ell+1)\ell}$ for $\ell=1,\dots,N-1$. The electric fields must thus linearly increase as a function of the layer index, $\ell$.

\subsection{Electric and magnetic dipoles}
Let us now turn to the in-plane sheet currents induced by the external fields. These currents give rise to electric and magnetic moments, and the total current density can be related to the electric polarization by
\begin{align}
-\partial_t\p&=\J_{\mathrm{tot}} =\sum_{\ell=1}^N\J_\ell\;.
\end{align}
Furthermore, in Ref. \onlinecite{Margetis24}, we outlined that the sheet current densities can be written as
\begin{equation}\label{eq:current-deviations}
\J_\ell = \frac{\J_{\mathrm{tot}}}{N} + \j_\ell\;,
\end{equation} 
where $\J_{\mathrm{tot}}=\sum_\ell \J_\ell$ denotes the total current and $\j_\ell$ the deviation from the average. Since $ \sum_\ell \j_\ell = 0$, each $ \j_\ell$ can be considered as the sum of  the magnetization currents associated with the regions above ($\m_{\ell}$) and below ($\m_{\ell-1}$) the layer $\ell$. For $\ell=1,\dots,N$ we get
\begin{equation}\label{eq:lower-js}
\m_{\ell-1} -\m_{\ell} = a (\e_z\times \j_{\ell})\,, 
\end{equation} 
with the constraints $ \sum_\ell \j_\ell= 0$ and $\m_0=\m_N=0$.
Equation \eqref{eq:lower-js} provides the set of magnetizations $\{\m_\ell\} $ associated with any set of currents $\{\j_\ell\} $, and vice versa. The total magnetic moment per unit area $\M=\sum_{\ell=1,\dots,N-1}\m_\ell$ is thus given by
\begin{align}\label{eq:mtotal}
\M=a\sum_{\ell=1}^{N}\frac{2\ell-N-1}{2}(\e_z\times\j_\ell)\;.
\end{align}
The in-plane magnetic response is therefore generated by layer-dependent current imbalances and encodes the spatial distribution of magnetization across the multilayer stack.

\section{In-plane magnetic response}
\label{sec:magnetic_response}
We now turn our attention to the in-plane magnetic response of alternating-twist multilayer systems. As shown in Sec.~\ref{sec:MagneticMoment}, an in-plane magnetic field 
is generated by a layer-discriminated electric field of the form
\begin{align}
\label{MagneticField}
{\bf E}_B^N=\E_0\sum_{\ell=1}^{N}\frac{2\ell-N-1}{2}\e_\ell\;,
\end{align}
where $\e_\ell$ denotes the unit vector associated with layer $\ell$ in the layer-space basis. The magnitude $B=|\B|$ of the magnetic field is thus given by the relation $i\omega a B=E_0$, where $E_0=|\E_0|$. 

In the remainder of this section, we apply the general formalism to the tetralayer and pentalayer systems, which illustrate the behavior of even- and odd-layer alternating-twist multilayers, respectively. For completeness, we also briefly review the trilayer case, derived in our previous work \cite{Margetis24}. For a more compact notation, we denote the in-plane current operator by two layer indices, $\J_{\ell}\to\j_{\ell\ell}$, which allows us to include ``vertical'' current densities $\j_{\ell\ell'}$ with $\ell\neq\ell'$; see Appendix~\ref{app:transformation}. Since the magnetic response is always parallel or antiparallel to the applied magnetic field, we suppress the boldface vector notation whenever 
no ambiguity arises.

\subsection{Trilayer response}
Following Ref. \cite{Margetis24}, the total magnetic moment per unit area for $N=3$ reads
\begin{align}
\M=2i\omega a^2(\sigma_0^{11}-\sigma_0^{13})\B\;.
\end{align}
After applying the unitary transformation, the counterflow response between the first and third layers is entirely determined by the cross term
$\sigma_c=2(\sigma_0^{11}-\sigma_0^{13})$, which can be written as
\begin{align}
\sigma_{c}&=\langle\langle  \bar j_{13} \bar j_{31}+\bar j_{31} \bar j_{13}\rangle\rangle\;,
\end{align}
using the transformed current operators $\bar{j}_{\ell\ell'}$; see Ref. \cite{Margetis24} and Appendix \ref{app:transformation}. The total susceptibility $\chi$ is defined through $\M=\chi\B$, which yields
\begin{align}
\chi=i\omega a^2\sigma_{c}.
\end{align}
Because of the kinematic constraints arising from the large mismatch between the Fermi velocities of the effective TBG and the decoupled SLG bands, the contribution $\sigma_c$ is expected to be negligibly small. It was shown in Ref.~\cite{Margetis24} that $\sigma_c$ is at most comparable to the corresponding atomistic (lattice) contribution in SLG.

\subsection{Tetralayer response}
By applying the electric field of Eq. (\ref{MagneticField}) to the system with $N=4$ layers, we obtain the following relations for the currents:
\begin{align}
j_{44}&=\frac{E_0}{2}\left[3(\sigma^{11}_0-\sigma^{14}_0)+(\sigma^{12}_0-\sigma^{13}_0)\right]=-j_{11}\;,\\
j_{33}&=\frac{E_0}{2}\left[3(\sigma^{12}_0-\sigma^{13}_0)+(\sigma^{22}_0-\sigma^{23}_0)\right]=-j_{22}\;,
\end{align}
where the conductivities have been reduced to the independent response functions dictated by the symmetry structure of Eq.~(\ref{eq:Sigma4}). Given that $i\omega a B=E_0$, where $B=|\B|$, and $\m_\ell=\chi_\ell\B$, we obtain
\begin{align}
\chi_1&=i\omega\frac{a^2}{2}\left[3(\sigma^{11}_0-\sigma^{14}_0)+(\sigma^{12}_0-\sigma^{13}_0)\right]=\chi_3\;,\\
\chi_2
&= i\omega\frac{a^2}{2}\Bigl[
3(\sigma^{11}_0-\sigma^{14}_0)
+4(\sigma^{12}_0-\sigma^{13}_0)
\nonumber\\
&\hspace{3.9em}
+(\sigma^{22}_0-\sigma^{23}_0)
\Bigr]\;.
\end{align}

After transforming the tetralayer into two effective TBG systems, the layer-resolved conductivities can be rewritten in the corresponding basis; see Appendix~\ref{app:tetra-currents}. Using the notation of Ref. \onlinecite{Stauber18}, we define
\begin{align}
\sigma_{0}^1
&=\langle\langle  \bar j_{11} \bar j_{11}\rangle\rangle
=\langle\langle  \bar j_{22} \bar j_{22}\rangle\rangle\;,
\label{eq:sigma01}\\
\sigma_{0}^2
&=\langle\langle  \bar j_{33} \bar j_{33}\rangle\rangle
=\langle\langle  \bar j_{44} \bar j_{44}\rangle\rangle\;,
\label{eq:sigma02}\\
\sigma_{1}^1
&=\langle\langle  \bar j_{11} \bar j_{22}\rangle\rangle
=\langle\langle  \bar j_{22} \bar j_{11}\rangle\rangle\;,
\label{eq:sigma11}\\
\sigma_{1}^2
&=\langle\langle  \bar j_{33} \bar j_{44}\rangle\rangle
=\langle\langle  \bar j_{44} \bar j_{33}\rangle\rangle\;.
\label{eq:sigma12}
\end{align}
Additionally, we have to introduce the responses that couple the two effective bilayer systems as follows:
\begin{align}
\sigma_{c}^1&=\langle\langle  \bar j_{13} \bar j_{31}+\bar j_{31} \bar j_{13}\rangle\rangle=\langle\langle  \bar j_{24} \bar j_{42}+\bar j_{42} \bar j_{24}\rangle\rangle\;,
\label{eq:sigma1c}\\
\sigma_{c}^2&=\langle\langle  \bar j_{13} \bar j_{42}+\bar j_{31} \bar j_{24}\rangle\rangle=\langle\langle  \bar j_{24} \bar j_{31}+\bar j_{42} \bar j_{13}\rangle\rangle\;.
\label{eq:sigma2c}
\end{align}
The final expressions for the conductivities $\sigma_0^{\ell\ell^\prime}$ can be found in Appendix~\ref{app:tetra-currents}. The response functions can now be written as
\begin{align}
\chi_1 &= i \omega \frac{a^2}{10} \Bigl[
(3\varphi^{-2}-1)(\sigma_0^1-\sigma_1^1) 
\nonumber\\
&\quad + (3\varphi^{2}-1)(\sigma_0^2-\sigma_1^2) 
+ 4(\sigma_c^1+\sigma_c^2)
\Bigr],\\[0.5em]
\chi_2 &= i \omega \frac{a^2}{10} \Bigl[
(3\varphi^{-2}-4+\varphi^{2})(\sigma_0^1-\sigma_1^1) 
\nonumber\\
&\quad + (3\varphi^{2}-4+\varphi^{-2})(\sigma_0^2-\sigma_1^2) + 8(\sigma_c^1+\sigma_c^2)\Bigr],
\end{align}
where $\varphi=(1+\sqrt{5})/2$. 

\subsubsection{System response near the first magic angle}
\label{sssec:1st-magang}
Since the magnetic response depends on the effective magic angle, $\theta^N_{k,m}$, we explicitly indicate the corresponding index $k$ throughout the remainder of this paper. We begin by considering the tetralayer system near the first effective magic angle, $\theta^4_{1,m}=\varphi\theta_m$.

The renormalized Fermi velocities of the two effective systems, $v_{\rm{F}}^j$, are very different at the $K$-point, $v_{\rm{F}}^2\gg v_{\rm{F}}^1$. The coupling term $\sigma_c^1+\sigma_c^2$ is thus expected to be relatively small because of the restricted phase space. The counterflow term $\sigma_0^2-\sigma_1^2$ also becomes negligible since the second effective bilayer is away from its magic angle and does not exhibit a flat-band enhancement. With respect to the susceptibility of the TBG, $\chi_{\mathrm{TBG}}=i\omega(a^2/2)(\sigma_0-\sigma_1)$, we arrive at
\begin{align}
\frac{\chi_1}{\chi_{\mathrm{TBG}}}(k=1)&=\frac{\varphi^{-4}}{5}\approx0.03\;,\\
\frac{\chi_2}{\chi_{\mathrm{TBG}}}(k=1)&=-\frac{\varphi^{-3}}{5}\approx-0.05\;.
\end{align}
By adding the susceptibilities of the three magnetizations, $\chi=2\chi_1+\chi_2$, we obtain
\begin{align}
\label{AverageSusc}
\frac{\chi}{\chi_{\mathrm{TBG}}}(k=1)=\frac{\varphi^{-6}}{5}\approx0.01\;.
\end{align}
One can see that the orbital in-plane magnetic response is negligible at the larger magic angle.

\subsubsection{System response near the second magic angle}
\label{sssec:2nd-magang}
The analysis proceeds analogously to the previous case (Sec.~\ref{sssec:1st-magang}) when the second effective bilayer system is tuned close to its magic angle, $\theta^4_{2,m}=\varphi^{-1}\theta_m$. The renormalized Fermi velocities at the $K$ point are strongly different, but now with $v_{\rm F}^1 \gg v_{\rm F}^2$. The coupling term $\sigma_c^1+\sigma_c^2$ is once again expected to be small because of the restricted phase space. The counterflow contribution $\sigma_0-\sigma_1$ of the first system can likewise be neglected, i.e., $\lvert \sigma_0^1 - \sigma_1^1 \rvert \ll \lvert \sigma_0^2 - \sigma_1^2 \rvert$. For the susceptibility relative to $\chi_{\mathrm{TBG}}$, we obtain
\begin{align}
\frac{\chi_1}{\chi_{\mathrm{TBG}}}(k=2)&=\frac{\varphi^{4}}{5}\approx1.4\;,\\
\frac{\chi_2}{\chi_{\mathrm{TBG}}}(k=2)&=\frac{\varphi^{3}}{5}\approx0.8\;.
\end{align}
By adding the susceptibilities of the three magnetizations according to $\chi=2\chi_1+\chi_2$, we finally get
\begin{align}
\label{AverageSusc2}
\frac{\chi}{\chi_{\mathrm{TBG}}}(k=2)=\frac{\varphi^{6}}{5}\approx3.6\;.
\end{align}
The total susceptibility at $\theta^4_{2,m}=\varphi^{-1}\theta_m$ is 
therefore larger by a factor $\varphi^{12}\approx322$ compared with its 
value at $\theta^4_{1,m}=\varphi\theta_m$.

\vspace{2em}
The results for the two tetralayer magic angles reveal a more general structure. In the flat-band regime near the $k$-th magic angle, the relation between the outermost interlayer susceptibility $\chi_1(k)=\chi_3(k)$ and the total susceptibility $\chi(k)$ can be 
directly written in terms of the corresponding scaling factor $\beta^4_k$ as
\begin{align}
\label{ChiBeta}
\chi_1(k)=\left(\beta^4_k\right)^2\chi(k),
\end{align}
where the superscript $N=4$ was omitted from the susceptibilities for simplicity, as throughout this section. Interestingly, this relation appears to be a universal property of alternating-twist multilayers with even $N$. In Appendix~C, we show that the corresponding generalized relation is also satisfied for the hexalayer and octalayer systems.

\subsection{Pentalayer response}
By applying the electric field of Eq.~(\ref{MagneticField}) to the system with
$N=5$ layers, while bearing in mind the symmetries of Eq.~(\ref{eq:Sigma5}), we obtain the following expressions for the layer currents $\J_{\ell}\to\j_{\ell\ell}$:
\begin{align}
j_{55}&=\left[2(\sigma^{11}_0-\sigma^{15}_0)+(\sigma^{12}_0-\sigma^{14}_0)\right]E_0=-j_{11}\;,\\
j_{44}&=\left[2(\sigma^{12}_0-\sigma^{14}_0)+(\sigma^{22}_0-\sigma^{24}_0)\right]E_0=-j_{22}\;,
\end{align} 
and $j_{33}=0$. In view of the relation $i\omega a B=E_0$, the susceptibilities read
\begin{align}
\chi_1 &= i \omega a^2 \Bigl[
2(\sigma^{11}_0-\sigma^{15}_0) + (\sigma^{12}_0-\sigma^{14}_0)
\Bigr] = \chi_4, \\[0.5em]
\chi_2 &= i \omega a^2 \Bigl[
2(\sigma^{11}_0-\sigma^{15}_0) + 3(\sigma^{12}_0-\sigma^{14}_0)
\nonumber\\
&\quad + (\sigma^{22}_0-\sigma^{24}_0)
\Bigr] = \chi_3.
\end{align}
After transforming the pentalayer system into two effective TBG systems and one decoupled effective single-layer, we can rewrite the layer-resolved conductivities in the corresponding basis (see Appendix~\ref{app:penta-currents}). It follows that the susceptibilities depend only on the cross terms and are given by
\begin{align}
\chi_1&=i\omega\frac{a^2}{6}\left[2\sigma_c^1+\sqrt{3}\sigma_c^2+4\sigma_c^4\right]\;,\\
\chi_2&=i\omega\frac{a^2}{6}\left[5\sigma_c^1+3\sqrt{3}\sigma_c^2+4\sigma_c^4\right]\;.
\end{align}
This confirms that the in-plane orbital magnetic response of the pentalayer is negligible, relying on the phase-space argument associated with the very different Fermi velocities of the effective TBG and SLG sectors. This behavior closely parallels that of the trilayer system, where the susceptibility also arises solely from cross terms. More generally, we find that the same mechanism applies to all alternating-twist multilayers with an odd number of layers, leading to a negligible in-plane orbital magnetic response.

\subsection{Discussion}
\label{sec:discussion}

For alternating-twist multilayers with an odd number of layers, we find a 
negligibly small in-plane orbital magnetic response, as shown for the trilayer and pentalayer systems. This suppression originates from the fact that the response is governed by cross terms between effective subsystems with very different Fermi velocities, which strongly restrict the available phase space.

\begin{figure}[!b]
\vspace{0.2cm}
\includegraphics[scale=0.38]{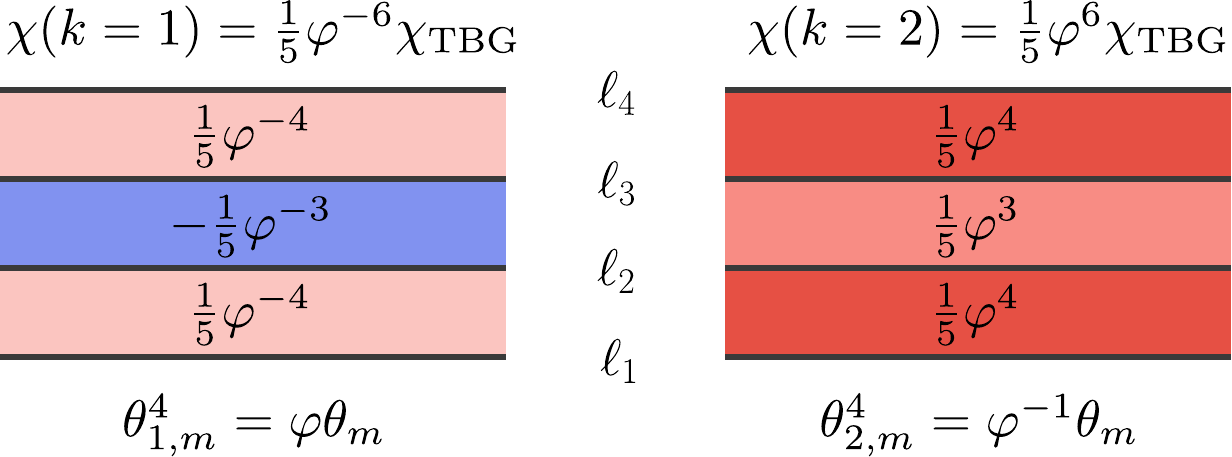}

\vspace{0.5cm}

\includegraphics[scale=0.57]{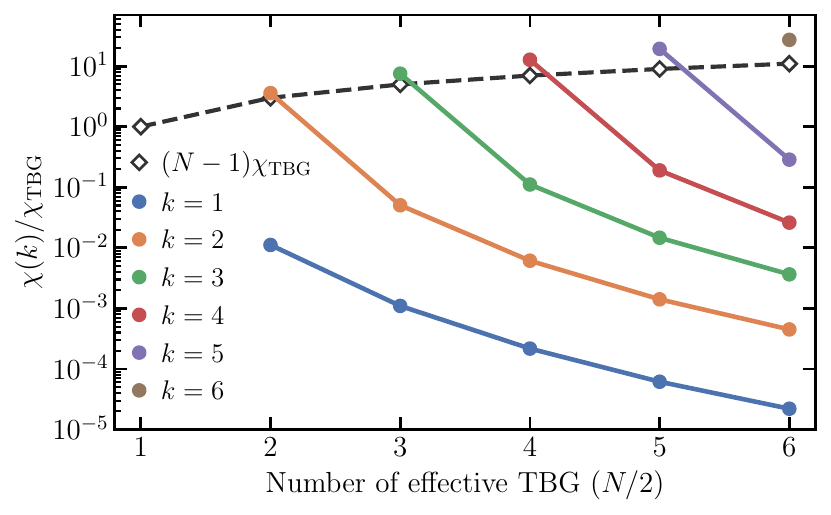}
\caption{
Upper panel: Schematic of the susceptibility profile induced by an in-plane magnetic field in a tetralayer system near the first (left) and second (right) magic angles. The interlayer susceptibility is given in units of $\chi_{\rm TBG}$. The color scale indicates the sign and relative magnitude of the magnetization in each interlayer region.
Lower panel: Total susceptibility $\chi(k)$, in units of $\chi_{\rm TBG}$, as a function of the number of effective TBG subsystems, $N/2$, for even-layer alternating-twist multilayers. The colored curves connect the values of $\chi(k)$ evaluated in the flat-band regime at the effective magic angles of the same order $k$, while the dashed line represents the linear scaling $(N-1)\chi_{\rm TBG}$, corresponding to the total number of interlayer spaces in an $N$-layer system.}
\label{fig:Magnetization}
\end{figure}

For even-layer systems, the response strongly depends on the particular magic angle within the hierarchy. In the tetralayer case, at the larger magic angle associated with $\varphi$, the total orbital susceptibility is reduced by approximately two orders of magnitude compared to magic-angle TBG. Moreover, the magnetization changes sign within the tetralayer, as illustrated in Fig.~\ref{fig:Magnetization} (upper left panel), implying the existence of a region between layers 2 and 3 where the local magnetization vanishes. This behavior is in stark contrast to the response at the smaller magic angle associated with $\varphi^{-1}$; see Fig.~\ref{fig:Magnetization} (upper right panel). In this case, the magnetization has the same sign throughout the system, leading to a strong orbital susceptibility that even exceeds the TBG value. Consequently, the measured magnetic response is expected to be dominated by the orbital contribution, with the spin susceptibility appearing only as a relatively small correction.

Even-layer alternating-twist multilayers exhibit additional universal properties. In the flat-band regime, the susceptibility associated with 
the outermost interlayer space is sufficient to determine the total orbital response of the system through the relation
\begin{align}
\chi_{\{1,N-1\}}^N(k)=\lambda_k^N\chi^N(k),
\end{align}
which generalizes Eq.~(\ref{ChiBeta}) to arbitrary even values of $N$. Here, $\lambda^N_k=(\beta^N_k)^2$ is the eigenvalue associated with 
each effective magic angle in the unitary transformation \cite{Khalaf19}. Moreover, the first magic angle (corresponding to the largest effective 
twist angle) always exhibits the smallest in-plane orbital susceptibility within the hierarchy, see Fig.~\ref{fig:Magnetization} (lower panel). Remarkably, this suppression becomes stronger as 
the number of layers increases, leading to a progressively smaller total susceptibility at the first magic angle for larger even-layer systems. Interestingly, for the finite even-layer systems studied in this work, only the smallest magic angle of each system ($k=N/2$) exhibits an enhanced magnetic response that exceeds the corresponding response of TBG. However, these angles are always below $1^\circ$, making their experimental realization increasingly challenging.

Our results demonstrate that different effective magic angles within the same alternating-twist even-layer system can exhibit qualitatively distinct magnetic responses. This has direct consequences for the interpretation of in-plane critical magnetic fields, since the measured response contains both spin and orbital contributions. In Sec.~\ref{sec:Maki}, we quantify this effect through the in-plane Maki parameter.

\section{In-plane Maki parameter}
\label{sec:Maki}
So far, we have implicitly assumed the multilayer system to be in the normal state. We now discuss the implications of our results for the superconducting phase. In this case, the ``contact term" vanishes independently of the filling factor due to the absence of a Fermi surface, i.e., as a consequence of the opening of a superconducting gap. Consequently, the equilibrium and Drude-like responses are equivalent \cite{Scalapino93}, and the in-plane magnetic response becomes diamagnetic at sufficiently large doping \cite{Stauber18,Stauber18b}. Since the same diamagnetic contribution is also present in the normal state \cite{Clogston62}, the dominant difference between the normal and superconducting phases arises from the Fermi-surface contribution. For dopings within the flat-band regime, the overall magnitude of this contribution is still set by the equilibrium response at $\mu=0$, $\chi_{\mathrm{TBG}}$, calculated in Sec.~\ref{sec:magnetic_response}. Our previous results can therefore be used to assess the possibility of extracting the spin susceptibility of Cooper pairs, $\chi_{\mathrm{Cooper}}$, in typical experiments, depending on whether the condition $\chi_{\mathrm{TBG}} \ll \chi_{\mathrm{Cooper}}$ is satisfied. For completeness, we now explicitly discuss the equilibrium response and the associated Fermi-surface contribution.

\subsection{Pauli limit and corrections}
The Pauli limit is related to the magnetic field that converts the superconducting phase into the normal state. It is usually assumed that the only contribution to the magnetic susceptibility is given by the spin response. This is obtained \cite{Clogston62} by equating the energies of the normal and superconducting phases in a magnetic field, in the absence of any Meissner effect. 
For an in-plane magnetic field applied to a planar system, the energy balance reads
\begin{align}\label{eq:DeltaF1}
\mathcal{F}_{\mathrm N}- \frac{1}{2} \chi^{\mathrm N}_{\rm{P}} B_{\rm{P}}^2  =
\mathcal{F}_{\mathrm S}- \frac{1}{2} \chi^{\mathrm S}_{\rm{P}} B_{\rm{P}}^2  
,\end{align}
where $\mathcal{F}_{\mathrm N,\mathrm S}$ is the free energy of the normal (N) or the  superconducting (S) phase, and $\chi^{\mathrm N,\mathrm S}_{\rm{P}} $ is the corresponding spin (Pauli) magnetic susceptibility. Under the stated assumptions, only the normal phase has spin susceptibility: $ \chi^{\mathrm N}_{\rm{P}} - \chi^{\mathrm S}_{\rm{P}} =\chi^{\mathrm N}_{\rm{P}}  = \chi_{\rm{P}} $, the standard Pauli spin susceptibility of the normal metal, $\chi_{\rm{P}} $. Thus, we have
\begin{align}\label{eq:DeltaF11}
\mathcal{F}_{\mathrm N}-\mathcal{F}_{\mathrm S}=\frac{1}{2} \chi_{\rm{P}} B_{\rm{P}}^2 
.\end{align}
Therefore, for the singlet Cooper pairs of the standard BCS theory, one obtains $B_{\rm{P}} = 1.86 T_{\rm{c}}$ (where $B_{\rm{P}}$ is in tesla if $T_{\rm{c}}$ is in kelvin).

The measurement of the critical magnetic field is not directly linked to the spin susceptibility if there is also an orbital contribution to the magnetic susceptibility, which is different for the normal and superconducting states. Accordingly, Eq.~(\ref{eq:DeltaF1}) should be amended to account for the critical magnetic field, $B_{\rm{c}}$, associated with the free-energy difference between the superconducting and normal phases:
\begin{align}\label{eq:DeltaF2}
\mathcal{F}_{\mathrm N}-\mathcal{F}_{\mathrm S}=\frac{1}{2} (\chi_{\rm{P}} + \Delta \chi_{\rm{orb}}) B_{\rm{c}}^2
,\end{align}
with $ \Delta \chi_{\rm{orb}} =\chi^{\mathrm N}_{\rm{orb}} - \chi^{\mathrm S}_{\rm{orb}}$, the difference between the orbital susceptibility of normal and superconducting phases. As discussed above, this difference arises from the Fermi-surface contribution, which we now discuss in detail.

In general, the equilibrium response of the quantity $A$ to a perturbation $V=-\lambda C$, to linear order in the parameter $\lambda$, is given by the susceptibility 
\begin{align} \label{eq:chiAC}
\chi_{AC}&=-\langle\langle A,C\rangle\rangle\\\notag
&+\frac{1}{S}\sum_{\k,n}\langle \k,n|A|\k,n\rangle\langle \k,n|C|\k,n\rangle\delta(\epsilon_{\rm{F}}-\epsilon_{\k,n})\;,
\end{align}
where $\k,n$ denote the Bloch momentum and the band number, respectively, $\epsilon_{\rm{F}}$ is the Fermi energy, and $S$ is the sample surface. The quantity $\chi_{AC}$ corresponds to the equilibrium ($\omega=0$, $\mathbf q\to 0$) response whereas  $-\langle\langle A,C\rangle\rangle$  denotes the Kubo-like response $(\mathbf q=0,\, \omega\to 0)$, already considered in this paper (see, for instance,  Eq.~(\ref{eq:resp-func}), where the current operators play the roles of $A$ and $C$). In the present case, the observable is the in-plane magnetic moment, such that $A=C=m_x$, and the perturbation parameter is the external magnetic field, $\lambda=B_x$.

In the superconducting phase, there is no Fermi surface and the equilibrium function is equal to the response from the Kubo formula \cite{Scalapino93}.  The latter, $\langle\langle A,C\rangle\rangle$ in Eq. (\ref{eq:chiAC}), is the same in the normal and superconducting phases up to corrections of order $\Delta/W $, the superconducting order parameter over the bandwidth, which we neglect. Therefore, only the Fermi surface term survives in Eq. (\ref{eq:chiAC}), as the dominant difference between superconducting and normal phases. In the remainder of this section, we estimate this contribution for the bilayer, tetralayer, and odd-layer alternating-twist systems.

\subsection{Bilayer systems}
As discussed above, the difference between the orbital susceptibilities of the normal and superconducting phases corresponds to the last term of Eq. (\ref{eq:chiAC}) with $A=C=m_x$, the in-plane orbital magnetic moment. For bilayer systems (BL), this term can be written as 
\begin{align}
\label{SusceptibilityOrb2}
\Delta \chi_{\rm{orb}}^{\rm{TBG}}= \frac{a^2}{2}D_{\rm{mag}}\;,
\end{align}
with 
\begin{align}
\label{MagneticDrude}
D_{\rm{mag}} &=  \frac{g_se^2}{S}\sum_{\k,n}\langle \k,n|j_{11}|\k,n\rangle\\&\times \notag\langle \k,n|j_{11}-j_{22}|\k,n\rangle\delta(\epsilon_{\rm{F}}-\epsilon_{\k,n})\;.
\end{align}
It follows that $D_{\rm{mag}}\geq0$, since $2\langle \k,n|j_{11}|\k,n\rangle\langle \k,n|j_{11}-j_{22}|\k,n\rangle=\langle \k,n|j_{11}-j_{22}|\k,n\rangle\langle \k,n|j_{11}-j_{22}|\k,n\rangle$.

Equation (\ref{eq:DeltaF2}) can now be written as 
\begin{align}\label{eq:DeltaF_maki}
\mathcal{F}_{\mathrm N}-\mathcal{F}_{\mathrm S}=\frac{1}{2} \chi_{\rm{P}} (1 + \alpha_{\rm{M}}) B_{\rm{c}}^2,
\end{align}
where we have introduced the {\it in-plane} Maki parameter (not to be confused with the original {\it out-of-plane} Maki parameter \cite{Maki66}),

\begin{figure}[!t]
\vspace{2mm}
\hspace{-2mm}
\includegraphics[scale=0.23]{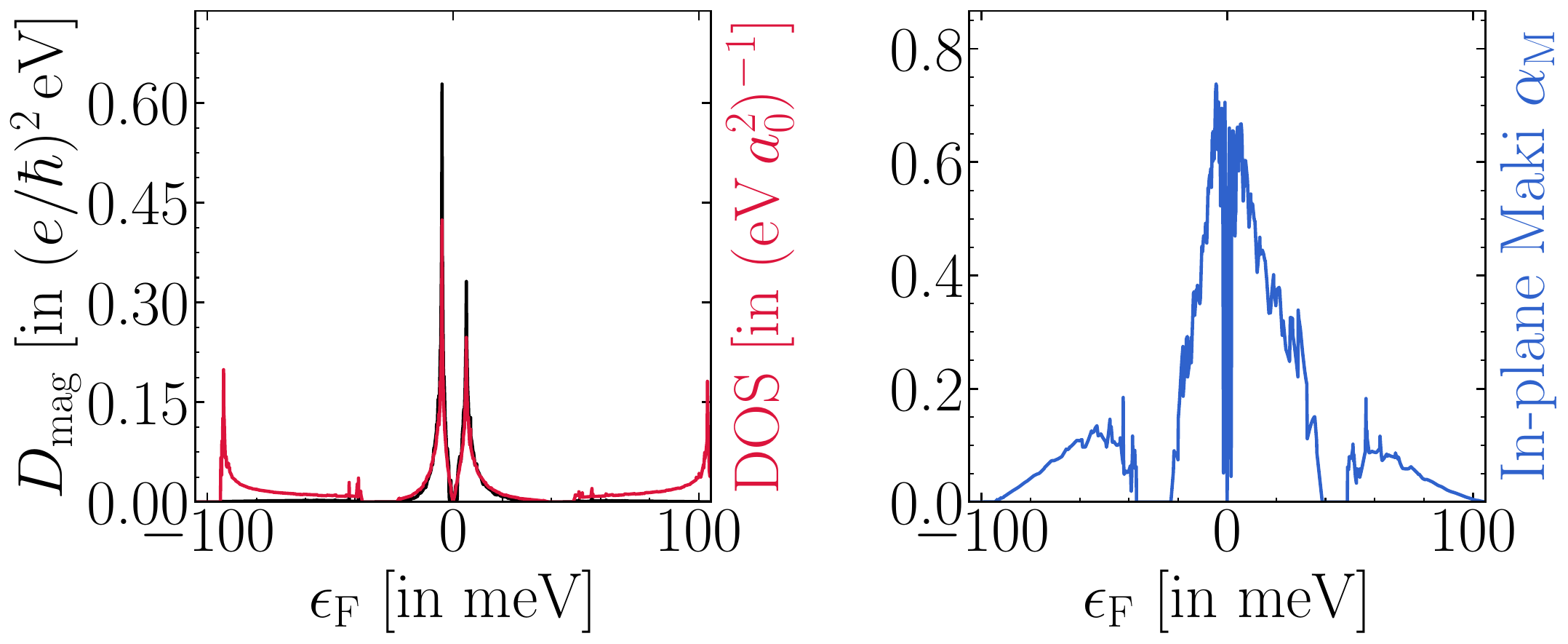}

\vspace{2mm}

\includegraphics[scale=0.23]{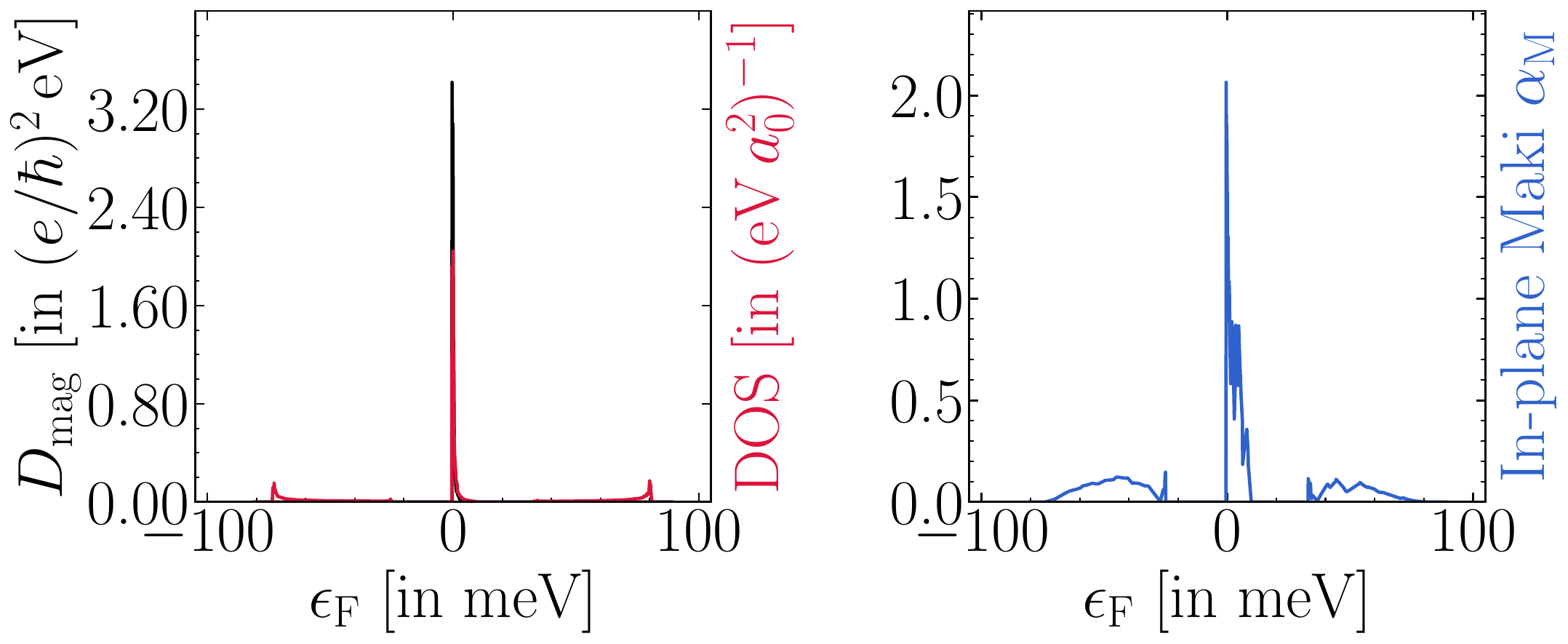}
\caption{Left: The orbital magnetic contribution of Eq. (\ref{SusceptibilityOrb2}) (black) and the density of states (DOS) of twisted bilayer graphene obtained from the non-interacting tight-binding model at a twist angle $\theta=1.25^\circ$ (upper panel)  and $\theta_m=1.05^\circ$ (lower panel), using the parameters of Ref. \cite{Sanchez24}. Right: The in-plane Maki parameter as defined in Eq. (\ref{Maki}).}
\label{fig:Drude}
\end{figure}

\begin{align}
 \alpha_{\rm{M}} =\frac{\Delta \chi_{\rm{orb}}}{\chi_{\rm{P}}},
\end{align}
as a measure of the  orbital correction to the usual Pauli limit. Combining Eqs.~(\ref{eq:DeltaF11}) and (\ref{eq:DeltaF_maki}), the hypothetical Pauli-limited magnetic field $B_{\rm P}$, which only couples to the spin susceptibility, can be related to the experimentally 
measured critical field $B_{\rm c}$ through
\begin{align}
B_{\rm P}=\sqrt{1+\alpha_{\rm M}}\,B_{\rm c}.
\end{align}
When discussing Pauli-limit violation, the relevant quantity is thus not the measured critical field itself, but rather the effective 
Pauli field after subtracting the orbital contribution since a large in-plane orbital susceptibility can strongly renormalize the measured critical magnetic field. Accordingly, agreement with or deviations from the standard Pauli limit in quasi-two-dimensional systems should not be regarded as evidence \textit{per se} for conventional or unconventional pairing symmetry, respectively, without accounting for the orbital contribution.

By using the Pauli susceptibility $\chi_{\rm{P}}=2\mu_{\rm{B}}^2\rho(\epsilon_{\rm{F}})$ with the Bohr magneton $\mu_{\rm{B}}=e\hbar/2m_e$ and the density of states (DOS) per spin channel, $
\rho(\epsilon_{\rm{F}})=\frac{1}{S}\sum_{\k,n}\delta(\epsilon_{\rm{F}}-\epsilon_{\k,n})\;,$ we obtain
 \begin{align}
 \label{Maki}
\alpha_{\rm{M}}=\left(\frac{aa_0 \rm{eV}}{\alpha a_{\rm{B}}\hbar c}\right)^2\frac{\tilde{D}_{\rm{mag}}}{\tilde{\rho}}\;=1.2\frac{\tilde{D}_{\rm{mag}}}{\tilde{\rho}},
\end{align}
where $a=3.4\,\mathrm{\AA}$ is the interlayer distance, $a_0=2.46\,\mathrm{\AA}$ the lattice constant, $\alpha$ is the fine-structure constant ($\alpha=1/137$), $a_{\rm{B}}$ is the Bohr radius ($a_{\rm{B}}=0.529\,\mathrm{\AA}$) and $c$ is the speed of light. We further have $D_{\mathrm{mag}}=\tilde{D}_{\mathrm{mag}}(e/\hbar)^2 \rm{eV}$ and $\rho=\tilde{\rho}(a_0^2 \rm{eV})^{-1}$.

The left-hand side of Fig.~\ref{fig:Drude} shows $D_{\mathrm{mag}}$ (black) and the density of states of the TBG obtained from the non-interacting tight-binding model at the twist angles $\theta=1.25^\circ$ (upper panel) and $\theta_m=1.05^\circ$ (lower panel) using the parameters of Ref.~\cite{Sanchez24}. The right-hand side of Fig.~\ref{fig:Drude} shows the in-plane Maki parameter defined in Eq.~(\ref{Maki}). In the magic-angle regime, the Maki parameter is of the order of unity and can reach values up to 2 throughout the valence band. Hence, the critical in-plane magnetic field that breaks superconductivity is greatly modified from its standard Pauli (spin) value due to the orbital magnetic contribution.

Furthermore, one can define an effective magnetic moment for the Bloch electrons $\mu_{\rm{orb}}$ as Eq. (\ref{MagneticDrude}) is a Fermi surface property. This gives
\begin{align}
\mu_{\rm{orb}}=\alpha_{\rm{M}}\mu_{\rm{B}}\;.
\end{align}
Our numerical analysis of $\mu_{\rm{orb}}\approx2\mu_{\rm{B}}$ for valence-band electrons at the magic angle agrees well with the estimate of Ref. \cite{Antebi2022}. 

Let us finally remark that for a nematic state that also breaks time-reversal symmetry, a permanent magnetic in-plane moment can emerge as first predicted by Antebi \textit{et al.} \cite{Antebi2022}. Using the results of Ref.~\cite{Sanchez24}, we calculate the in-plane magnetic moments for the nematic states corresponding to the valley-polarized phase ($\nu=-2$, $U=4\,\mathrm{eV}$) and the intervalley-coherent phase ($\nu=2$, $U=0.5\,\mathrm{eV}$) as a function of the effective interaction strength $\alpha=e^2/(4\pi\epsilon_0\epsilon)$, expressed in units of $\mathrm{eV}\times a_0$. At a critical interaction strength $\alpha\sim0.3$, there is a phase transition to a $C_6$- and $C_3$-symmetric state, respectively, in which no magnetic moment can emerge.

\begin{figure}[h]
\hspace{-2mm}
\includegraphics[scale=0.35]{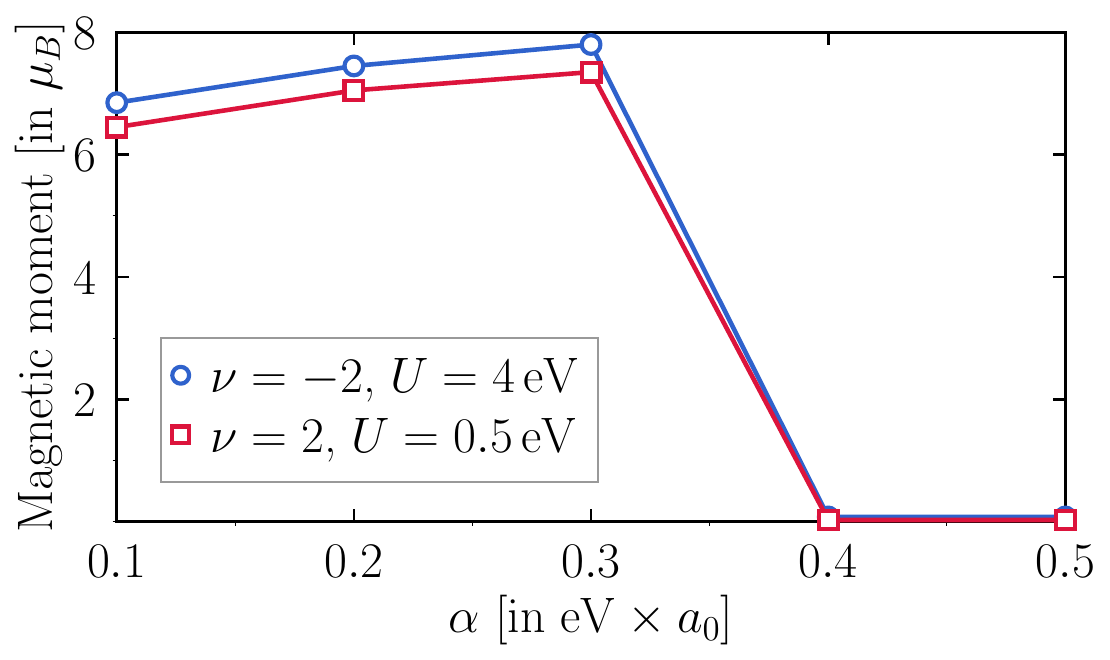}
\vspace{-2.5mm}
\caption{Emerging in-plane magnetic moment of TBG as a function of $\alpha=e^2/(4\pi\epsilon_0\epsilon)$ (in units of eV$\times a_0$) in the
nematic state for filling factors $\nu=-2$ (blue) and $\nu=2$ (red). The symmetry-broken state is
valley-polarized in the former case and Kramers intervalley coherent in the latter. In the strong-coupling regime ($\alpha\gtrsim0.3$), both
systems undergo a phase transition to a non-nematic state, leading to a vanishing in-plane magnetic moment; see Ref.~\cite{Sanchez24}.}
\label{fig:mm}
\end{figure}

The results are shown in Fig.~\ref{fig:mm} including the abrupt drop of the permanent magnetic in-plane moment at the phase transition. For the weak-coupling regime, the in-plane magnetic moment $\mu\sim8\mu_B$ can now be related to the Maki parameter $\alpha_M\sim8$. As in the first experiment on superconductivity in TBG, a critical field of $B_{\rm{P}}=1.86\,T_{\rm{c}}$ was measured \cite{Cao_2018unconv}, this value implies a Pauli-factor violation by a factor of 3, comparable to what is found for alternating-twist multilayers \cite{Park22}.

Our findings thus suggest a common pairing mechanism for alternating-twist multilayers, including the special case of TBG. We can further speculate on the pairing symmetry of the Cooper pairs. Assuming opposite-spin pairing, the spins should therefore be polarized perpendicular to the direction of the magnetization, so that the Cooper pairs are not affected asymmetrically by the in-plane magnetic moment. Residual Kane-Mele spin-orbit coupling \cite{Kane2005,Stauber23} may further lock the spins out of plane, yielding a situation closely resembling Ising superconductivity in proximitized bilayer graphene \cite{Zhang2023}.

\subsection{Tetralayer systems}
Following the above reasoning, for $N=4$ layers we obtain 
\begin{align}
\label{Delta-tetra}
\Delta \chi_{\rm orb}^{N=4}
&= \frac{a^2}{2}\frac{g_s e^2}{S}\sum_{\k,n}
\Bigl[
9\langle j_{11}\rangle(\langle j_{11}\rangle-\langle j_{44}\rangle) \notag\\
&\quad +6\langle j_{11}\rangle(\langle j_{22}\rangle-\langle j_{33}\rangle) \notag\\
&\quad +\langle j_{22}\rangle(\langle j_{22}\rangle-\langle j_{33}\rangle)
\Bigr]
\delta(\epsilon_{\rm F}-\epsilon_{\k,n})\; ,
\end{align}
where we have suppressed the dependence on the quantum numbers for simplicity.

The spectrum of the tetralayer system is the sum of the spectra of the two effective bilayer systems (TBG). Therefore, together with the transformation of the current operator, Eq.~\eqref{Delta-tetra} can be approximately mapped onto the orbital contribution of the bilayer systems, Eq. (\ref{SusceptibilityOrb2}). For this mapping, we assume large counterflow close to the magic angle, first discussed in Ref. \cite{bistritzer}, and only retain quadratic contributions. It is worthwhile noting that the cross terms now vanish identically.

By these approximations, we obtain the same expressions as in Sec.~\ref{sec:magnetic_response}, which read
\begin{align}
\frac{\Delta \chi_{\mathrm{orb}}^{N=4}}
{\Delta \chi_{\mathrm{orb}}^{\mathrm{TBG}}}
&=\frac{1}{5}(9\varphi^{-2}-6+\varphi^2)
=\frac{\varphi^{-6}}{5}\approx0.01,
&\theta\sim\theta_{1,m}^4,\\
\frac{\Delta \chi_{\mathrm{orb}}^{N=4}}
{\Delta \chi_{\mathrm{orb}}^{\mathrm{TBG}}}
&=\frac{1}{5}(9\varphi^{2}-6+\varphi^{-2})
=\frac{\varphi^{6}}{5}\approx3.6,
&\theta\sim\theta_{2,m}^4.
\end{align}
At hole doping, the in-plane Maki parameter changes from $\alpha_{\rm{M}}\approx0.02$ at the first magic angle to $\alpha_{\rm{M}}\approx7$ at the second. Accordingly, the standard interpretation of the Pauli limit applies only near the first magic angle. 

\subsection{Odd-layer systems}
For alternating-twist multilayers with an odd number of layers, the orbital susceptibility vanishes within the continuum description. After 
applying the unitary transformation, the in-plane magnetic response is determined only by cross terms coupling different effective subsystems. 
These contributions vanish due to particle conservation in the decoupled basis, leaving only negligible atomistic corrections beyond the continuum model. This explains why the violation of the Pauli limit in odd-layer systems can be directly associated with the spin susceptibility of the Cooper pairs.


\section{Summary}
\label{sec:conclusion}
In this paper, we have analytically studied the orbital in-plane magnetic response of alternating-twist graphene multilayers. Our approach is based on the unitary transformation introduced in Ref.~\cite{Khalaf19}, which maps an $N$-layer system onto a set of effective TBG subsystems, together with an additional decoupled SLG for odd values of $N$. This allows us to express the magnetic response of alternating-twist multilayers in terms of the corresponding effective TBG responses, substantially extending our previous analysis of the alternating-twist trilayer system \cite{Margetis24}. We have explicitly studied the tetralayer and pentalayer systems as representative examples of even- and odd-layer structures, respectively.

For systems with an odd number of layers, we find that the in-plane orbital magnetic response is negligibly small, since it is governed only by cross terms between effective subsystems with very different Fermi velocities.
In contrast, for a system with an even number of layers, specifically for $N=4$, a strong magnetic response could be expected. However, the behavior of the response at the two magic angles is markedly different. At the first magic angle, $\theta_{4,m}^1\approx1.70^\circ$, the magnetization is small, and a sign change in the susceptibility across the central layers leads to regions with vanishing magnetization around layers 2 and 3. At the second magic angle, $\theta_{4,m}^2\approx0.65^\circ$, the magnetization points uniformly in the same direction as the magnetic field and reaches a magnitude comparable to that of TBG.

For even-layer systems, we have found a universal hierarchy of orbital magnetic responses controlled by the effective magic angles generated by the unitary transformation. In the flat-band regime, the first magic angle always exhibits the weakest orbital response, with an increasing suppression for larger even-layer systems, whereas the smallest magic angles of each multilayer enhance the magnetic response. Another interesting finding is that the outermost interlayer space and the total susceptibility are linearly related by the eigenvalue associated with each magic angle.

We also introduced and studied the in-plane Maki parameter, $\alpha_{\rm{M}}$, as a measure of the relative importance of the orbital magnetic susceptibility in the modification of the standard Pauli (spin) limit of superconductivity. Notably, we obtained values up to 2 for $\alpha_{\rm{M}}$ at the magic angle for the TBG. In a symmetry broken state, we even find values up to 8, suggesting that twisted bilayer graphene shows a similar Pauli-limit violation as alternating-twist multilayers. We argued that this result can also be used for the tetralayer system, and obtained $\alpha_{\rm{M}}\approx0.02$ and $\alpha_{\rm{M}}\approx7$ for twist angles around the larger and smaller magic angle, respectively.

Our results motivate further studies on the electromagnetic response of alternating-twist multilayer graphene systems. First, the intrinsic chirality of tetralayers may give rise to nontrivial effects. Moreover, the investigation of superconductivity at the second effective magic angle would be of interest, since the enhanced orbital response can strongly modify the relation between the measured critical field and the underlying Pauli spin susceptibility.

\acknowledgments{
The authors thank E. Kaxiras, M. Luskin, and Z. Zhu for useful discussions.
I.V. was supported by Grant No. PREP2023-001135, funded by the Spanish Ministry of Science, Innovation and Universities through the State Research Agency (AEI).
The work of T.S. was supported by Grants No. PID2020-113164GB-I00 and No. PID2023-146461NB-I00, funded by the Spanish Ministry of Science, Innovation and Universities, and by the CSIC Research Platform on Quantum Technologies (PTI-001).
M.S.S. was supported by Grant No. PRE2021-097070, funded by the Spanish Ministry of Science, Innovation and Universities through the State Research Agency (AEI).
G.G.-S. acknowledges support from the Spanish Ministry of Science, Innovation and Universities through the ``María de Maeztu'' Programme for Units of Excellence in R\&D (Grant No. CEX2023-001316-M).
}

\appendix
\section{Decoupling the $N$-layer Hamiltonian}
\label{app:decoupling}

Following Ref.~\onlinecite{Khalaf19}, we consider the following Hamiltonian
\begin{align} 
H= 
\begin{bmatrix} 
H_{1} &V&\dots &\dots&0\\
V^\dagger&H_{2}&V&\dots&0\\
\vdots &\ddots&\ddots&\ddots&\vdots\\
0&\dots&V^\dagger&H_{N-1}&V\\
0 &\dots&\dots&V^\dagger & H_{N}
\end{bmatrix}.
\end{align} 
The $m\times m$-dimensional matrices $H_\ell$ and $V$ denote the Hamiltonian of layer $\ell$ and the interlayer coupling, respectively. 

Let us now set $H_{2\ell-1}=H_{(-\theta/2)}$ and $H_{2\ell}=H_{(\theta/2)}$ with $\ell=1,\dots, N/2$ and $N$ even. For $N$ odd, the decoupling procedure is equivalent, only with one additional decoupled single-layer Hamiltonian.

Rearranging the layers, we can write the Hamiltonian in the following way
\begin{align} 
H\to 
\begin{bmatrix} 
H_{(-\theta/2)}^{N/2} &W\\
W^\dagger&H_{(\theta/2)}^{N/2}
\end{bmatrix}\;,
\end{align} 
with $H_{(\pm\theta/2)}^{N/2}=H_{(\pm\theta/2)}\otimes\mathbf{1}_{N/2}$ and $W=V\otimes C^{N/2}$ with
\begin{align} 
C^{N/2}= 
\begin{bmatrix} 
1&0&\dots &0\\
1& \ddots&\ddots&0\\
\vdots &\ddots&\ddots&\vdots\\
0 &\dots&1 &1
\end{bmatrix}.
\end{align} 
We can now decompose the non-Hermitian matrix $C^{N/2}$ by its singular values, i.e., $C^{N/2}=A^{N/2}C^{N/2}_{s}\left(B^{N/2}\right)^{\dagger}$, where $C^{N/2}_{s}=\text{diag}(\beta_k^N)$ with $\lambda_k^{N}=(\beta_k^{N})^2$ the eigenvalues of $C^{N/2}(C^{N/2})^\dagger$. Furthermore, the columns of $A^{N/2}$ are the eigenvectors of $C^{N/2}(C^{N/2})^{\dagger}$ and the columns of $B^{N/2}$ are the eigenvectors of $(C^{N/2})^{\dagger}C^{N/2}$.

This decomposition now allows us to define the final unitary transformation $D=\text{diag}(\mathbf{1}_{m}\otimes A^{N/2},\mathbf{1}_{m}\otimes B^{N/2})$ to yield
\begin{align} 
H_s=D^\dagger HD= 
\begin{bmatrix} 
H_{(-\theta/2)}^{N/2} &W_s\\
W_s^\dagger&H_{(\theta/2)}^{N/2}
\end{bmatrix}\;,
\end{align}
where $W_s=V\otimes C^{N/2}_s$. Rearranging the Hamiltonian $H_s$, we thus arrive at a Hamiltonian that consists of the direct sum of twisted bilayer Hamiltonians with effective twist angles:
\begin{align} 
H\to
\begin{bmatrix}H^{\mathrm{TBG}}_{(\beta^{N}_{1})} & 0 & \dots & 0\\
0 & H^{\mathrm{TBG}}_{(\beta^{N}_{2})} & \ddots & 0\\
\vdots & \ddots & \ddots & 0\\
0 & \dots & 0 & H^{\mathrm{TBG}}_{(\beta^{N}_{N/2})}
\end{bmatrix},
\end{align}
where $H_{(\beta_k^N)}^{\mathrm{TBG}}$ denotes the $2m\times 2m$-dimensional Hamiltonian of the twisted bilayer with renormalized interlayer hopping amplitude $(t_k^N)_\perp=\beta_k^N t_\perp$ with $\beta_k^{N}=2\cos[\pi k/(N+1)],\; k=1,\dots,\lfloor N/2\rfloor$ \cite{Khalaf19}. For small initial twist angle $\theta$ and few layers, this is equivalent to the renormalization of the twist angle of the multilayer system, $\theta_{k,m}^{N}=\beta_k^{N}\theta_m$ \cite{bistritzer}. 
\section{Transformation of the current operators}
\label{app:transformation}
\subsection{Tetralayer transformation}
\label{app:tetra-currents}
For a system with four layers, the matrices that yield the singular value decomposition read
\begin{align} 
A=\frac{5^{-1/4}}{\sqrt{\varphi}}
\begin{bmatrix} 
1&-\varphi\\
\varphi&1
\end{bmatrix}\;,
B=\frac{5^{-1/4}}{\sqrt{\varphi}}
\begin{bmatrix} 
\varphi&-1\\
1&\varphi
\end{bmatrix}\;,
\end{align} 
where $\varphi=(1+\sqrt{5})/2$. Reordering the Hamiltonian again, we obtain the unitary matrix that block-diagonalizes the Hamiltonian, i.e.,
\begin{align}\label{eq:H_transform}
H\to T^\dagger H T=
\begin{pmatrix} 
H_{(\varphi)}^{\mathrm{TBG}}&0\\
0&H_{(\varphi^{-1})}^{\mathrm{TBG}}                                               
\end{pmatrix}\;,
\end{align}
where $H^{\mathrm{TBG}}_{(\beta^4_k)}$ is the Hamiltonian of the twisted bilayer with the two renormalized interlayer coupling amplitudes $(t_k^4)_\perp=\beta_k^{4}t_\perp$. The effective twist angle can therefore be approximated by $\theta^{4}_k=\beta_{k}^{4}\theta_m$, so that the first two magic angles read $\theta_{1,m}^{4}=\varphi\theta_m\approx1.70^\circ$ and $\theta_{2,m}^{4}=\varphi^{-1}\theta_m\approx0.65^\circ$, with $\theta_m=1.05^\circ$.

The unitary matrix is
\begin{align}\label{eq:matrix-T}
 T=\frac{5^{-1/4}}{\sqrt{\varphi}}
\begin{pmatrix} 
1&0&-\varphi&0\\
0&\varphi&0&-1\\
\varphi&0&1&0\\
0&1&0&\varphi                                               
\end{pmatrix}\;,
\end{align}
where each component is proportional to the $m\times m$-identity matrix, with $m=2$ in the case of an underlying continuum model or $m=2A_M$ in the case of a tight-binding model with commensurate twist angle $\theta=\arccos[1-1/(2A_M)]$, where $A_M=3M^3+3M+1$. This matrix relates the initial layer operators $c_\ell$ to the transformed layer operators $\bar c_\ell$:
\begin{align}
c_1&=\frac{5^{-1/4}}{\sqrt{\varphi}}(\bar c_1-\varphi \bar c_3)\;,\\
c_2&=\frac{5^{-1/4}}{\sqrt{\varphi}}(\varphi\bar c_2-\bar c_4)\;,\\
c_3&=\frac{5^{-1/4}}{\sqrt{\varphi}}(\varphi\bar c_1+\bar c_3)\;,\\
c_4&=\frac{5^{-1/4}}{\sqrt{\varphi}}(\bar c_2+\varphi\bar c_4)\;.
\end{align}
The layer current is now related to the bilinear combination of the layer operators $j_{\ell\ell}\propto c^\dagger_\ell c_\ell$. Note that we could have omitted the second layer index $j_{\ell\ell}\to j_\ell$ as we only discuss in-plane sheet current densities. This current can now be related to the transformed current $\bar j_{\ell\ell'}\propto \bar c^\dagger_\ell \bar c_{\ell'}$ as follows:
\begin{align}
j_{11}&=\frac{5^{-1/2}}{\varphi}[\bar j_{11}+\varphi^2\bar j_{33}-\varphi(\bar j_{13}+\bar j_{31})]\;,\\
j_{22}&=\frac{5^{-1/2}}{\varphi}[\varphi^2\bar j_{22}+\bar j_{44}-\varphi(\bar j_{24}+\bar j_{42})]\;,\\
j_{33}&=\frac{5^{-1/2}}{\varphi}[\varphi^2\bar j_{11}+\bar j_{33}+\varphi(\bar j_{13}+\bar j_{31})]\;,\\
j_{44}&=\frac{5^{-1/2}}{\varphi}[\bar j_{22}+\varphi^2\bar j_{44}+\varphi(\bar j_{24}+\bar j_{42})]\;.
\end{align}
There are six independent correlation functions, while the remaining ten can be obtained from the symmetries of the tetralayer conductivity tensor given in Eq.~(\ref{eq:Sigma4}). Writing these independent correlators explicitly in terms of the transformed current operators, we obtain:
\begin{align}
\langle\langle j_{11}j_{11}\rangle\rangle
&=\frac{1}{5\varphi^2}\Bigl[\langle\langle \bar j_{11}\bar j_{11}\rangle\rangle
+\varphi^4\langle\langle  \bar j_{33} \bar j_{33}\rangle\rangle \notag\\
&\qquad
+\varphi^2\langle\langle  \bar j_{13} \bar j_{31}+\bar j_{31} \bar j_{13}\rangle\rangle\Bigr]\;,\\[0.4em]
\langle\langle j_{11}j_{22}\rangle\rangle
&=\frac{1}{5\varphi^2}\Bigl[\varphi^2\langle\langle \bar j_{11}\bar j_{22}\rangle\rangle
+\varphi^2\langle\langle  \bar j_{33} \bar j_{44}\rangle\rangle \notag\\
&\qquad
+\varphi^2\langle\langle  \bar j_{13} \bar j_{42}+\bar j_{31} \bar j_{24}\rangle\rangle\Bigr]\;,\\[0.4em]
\langle\langle j_{11}j_{33}\rangle\rangle
&=\frac{1}{5\varphi^2}\Bigl[\varphi^2\langle\langle \bar j_{11}\bar j_{11}\rangle\rangle
+\varphi^2\langle\langle  \bar j_{33} \bar j_{33}\rangle\rangle \notag\\
&\qquad
-\varphi^2\langle\langle  \bar j_{13} \bar j_{31}+\bar j_{31} \bar j_{13}\rangle\rangle\Bigr]\;,\\[0.4em]
\langle\langle j_{11}j_{44}\rangle\rangle
&=\frac{1}{5\varphi^2}\Bigl[\langle\langle \bar j_{11}\bar j_{22}\rangle\rangle
+\varphi^4\langle\langle  \bar j_{33} \bar j_{44}\rangle\rangle \notag\\
&\qquad
-\varphi^2\langle\langle  \bar j_{13} \bar j_{42}+\bar j_{31} \bar j_{24}\rangle\rangle\Bigr]\;,\\[0.4em]
\langle\langle j_{22}j_{22}\rangle\rangle
&=\frac{1}{5\varphi^2}\Bigl[\varphi^4\langle\langle \bar j_{22}\bar j_{22}\rangle\rangle
+\langle\langle  \bar j_{44} \bar j_{44}\rangle\rangle \notag\\
&\qquad
+\varphi^2\langle\langle  \bar j_{24} \bar j_{42}+\bar j_{42} \bar j_{24}\rangle\rangle\Bigr]\;,\\[0.4em]
\langle\langle j_{22}j_{33}\rangle\rangle
&=\frac{1}{5\varphi^2}\Bigl[\varphi^4\langle\langle \bar j_{22}\bar j_{11}\rangle\rangle
+\langle\langle  \bar j_{44} \bar j_{33}\rangle\rangle \notag\\
&\qquad
-\varphi^2\langle\langle  \bar j_{24} \bar j_{31}+\bar j_{42} \bar j_{13}\rangle\rangle\Bigr]\;.
\end{align}
Using the notation introduced in Eqs.~(\ref{eq:sigma01})--(\ref{eq:sigma2c}), 
which makes the connection to the effective TBG subsystems explicit, the 
magnetic susceptibilities of the tetralayer system can be rewritten as:
\begin{align}
\sigma_0^{11}&=\frac{1}{5}\left[\varphi^{-2}\sigma_0^1+\varphi^2\sigma_0^2+\sigma_c^1\right]\;,\\
\sigma_0^{12}&=\frac{1}{5}\left[\sigma_1^1+\sigma_1^2+\sigma_c^2\right]\;,\\
\sigma_0^{13}&=\frac{1}{5}\left[\sigma_0^1+\sigma_0^2-\sigma_c^1\right]\;,\\
\sigma_0^{14}&=\frac{1}{5}\left[\varphi^{-2}\sigma_1^1+\varphi^2\sigma_1^2-\sigma_c^2\right]\;,\\
\sigma_0^{22}&=\frac{1}{5}\left[\varphi^2\sigma_0^1+\varphi^{-2}\sigma_0^2+\sigma_c^1\right]\;,\\
\sigma_0^{23}&=\frac{1}{5}\left[\varphi^2\sigma_1^1+\varphi^{-2}\sigma_1^2-\sigma_c^2\right]\;.
\end{align}
These expressions allow one to compute the magnetic susceptibilities of the tetralayer system.
\subsection{Pentalayer transformation}
\label{app:penta-currents}
Following Ref.~\onlinecite{Khalaf19}, we start from the unitary transformation that block-diagonalizes the Hamiltonian, i.e.
\begin{align}\label{eq:matrix-Tpenta}
H=T^\dagger H T=
\begin{pmatrix} 
H^{\mathrm{TBG}}_{(\sqrt{3})}&0&0\\
0&H^{\mathrm{TBG}}_{(1)}&0\\
0&0&H^{\mathrm{SLG}}                                              
\end{pmatrix}\;,
\end{align}
where $H^{\mathrm{TBG}}_{(\beta^5_k)}$ is the Hamiltonian of the TBG with the two renormalized interlayer coupling amplitudes $(t_k^N)_\perp=\beta_k^5t_\perp$ and $H^{\mathrm{SLG}}$ the Hamiltonian of a monolayer graphene. The effective twist angles are approximated by $\theta^{5}_{k,m}=\beta_{k}^{5}\theta_m$, giving $\theta_{1,m}^{5}=\sqrt3\theta_m\approx1.87^\circ$ and $\theta_{2,m}^{5}=\theta_m\approx1.05^\circ$, with $\theta_m$ the magic angle of TBG.

The original layer operators $c_\ell$ are related to the transformed operators $\bar c_\ell$ as:
\begin{align}
c_1&=\frac{1}{\sqrt{6}}(\bar c_1-\sqrt{3} \bar c_3+\sqrt{2}\bar c_5)\;,\\
c_2&=\frac{1}{\sqrt{2}}(\bar c_2-\bar c_4)\;,\\
c_3&=\frac{1}{\sqrt{3}}(\sqrt{2}\bar c_1-\bar c_5)\;,\\
c_4&=\frac{1}{\sqrt{2}}(\bar c_2+\bar c_4)\;,\\
c_5&=\frac{1}{\sqrt{6}}(\bar c_1+\sqrt{3} \bar c_3+\sqrt{2}\bar c_5)\;.
\end{align}
Proceeding as for the tetralayer system, the layer currents are expressed in terms of the transformed current operators $\bar j_{\ell\ell'}\propto \bar c^\dagger_\ell \bar c_{\ell'}$ through:

\begin{align}
j_{11}
&=\frac{1}{6}(
\bar j_{11}+3\bar j_{33}+2\bar j_{55}
-\sqrt{3}(\bar j_{13}+\bar j_{31}) \notag\\
&\qquad
+\sqrt{2}(\bar j_{15}+\bar j_{51})
-\sqrt{6}(\bar j_{35}+\bar j_{53})
)\;,\\[0.4em]
j_{22}
&=\frac{1}{2}(
\bar j_{22}+\bar j_{44}
-(\bar j_{24}+\bar j_{42})
)\;,\\[0.4em]
j_{33}
&=\frac{1}{3}(
2\bar j_{11}+\bar j_{55}
-\sqrt{2}(\bar j_{15}+\bar j_{51})
)\;,\\[0.4em]
j_{44}
&=\frac{1}{2}(
\bar j_{22}+\bar j_{44}
+(\bar j_{24}+\bar j_{42})
)\;,\\[0.4em]
j_{55}
&=\frac{1}{6}(
\bar j_{11}+3\bar j_{33}+2\bar j_{55}
+\sqrt{3}(\bar j_{13}+\bar j_{31}) \notag\\
&\qquad
+\sqrt{2}(\bar j_{15}+\bar j_{51})
+\sqrt{6}(\bar j_{35}+\bar j_{53})
)\;.
\end{align}
The nine independent correlation functions can be expressed in terms of these transformed current operators. Using the same notation introduced for 
the tetralayer system in Eqs.~(\ref{eq:sigma01})--(\ref{eq:sigma12}), the connection to the two effective TBG subsystems becomes explicit. In 
addition, we introduce $\sigma_{0}^3=\langle\langle \bar j_{55}\bar j_{55}\rangle\rangle$ to describe the conductivity of the decoupled single-layer graphene. The coupling terms between the different effective subsystems are defined as
\begin{align}
\sigma_{c}^1&=\langle\langle  \bar j_{13} \bar j_{31}+\bar j_{31} \bar j_{13}\rangle\rangle=\langle\langle  \bar j_{24} \bar j_{42}+\bar j_{42} \bar j_{24}\rangle\rangle\;,
\label{eq:sigma1cpenta}\\
\sigma_{c}^2&=\langle\langle  \bar j_{13} \bar j_{42}+\bar j_{31} \bar j_{24}\rangle\rangle=\langle\langle  \bar j_{24} \bar j_{31}+\bar j_{42} \bar j_{13}\rangle\rangle\;,
\label{eq:sigma2cpenta}\\
\sigma_{c}^3&=\langle\langle  \bar j_{15} \bar j_{51}+\bar j_{51} \bar j_{15}\rangle\rangle,\\
\sigma_{c}^4&=\langle\langle  \bar j_{35} \bar j_{53}+\bar j_{53} \bar j_{35}\rangle\rangle.
\end{align}
The independent layer-resolved conductivities are then given by:

\begin{align}
\sigma_0^{11}&=\frac{1}{36}\left[\sigma_0^1+9\sigma_0^2+4\sigma_0^3+3\sigma_c^1+2\sigma_c^3+6\sigma_c^4\right]\;,\\
\sigma_0^{12}&=\frac{1}{12}\left[\sigma_1^1+3\sigma_1^2+\sqrt{3}\sigma_c^2\right]\;,\\
\sigma_0^{13}&=\frac{1}{9}\left[\sigma_0^1+\sigma_0^3-\sigma_c^3\right]\;,\\
\sigma_0^{14}&=\frac{1}{12}\left[\sigma_1^1+3\sigma_1^2-\sqrt{3}\sigma_c^2\right]\;,\\
\sigma_0^{15}&=\frac{1}{36}\left[\sigma_0^1+9\sigma_0^2+4\sigma_0^3-3\sigma_c^1+2\sigma_c^3-6\sigma_c^4\right]\;,\\
\sigma_0^{22}&=\frac{1}{4}\left[\sigma_0^1+\sigma_0^2+\sigma_c^1\right]\;,\\
\sigma_0^{23}&=\frac{1}{3}\sigma_1^1\;,\\
\sigma_0^{24}&=\frac{1}{4}\left[\sigma_0^1+\sigma_0^2-\sigma_c^1\right]\;,\\
\sigma_0^{33}&=\frac{1}{9}\left[4\sigma_0^1+\sigma_0^3+2\sigma_c^3\right]\;
\end{align}
The remaining layer-resolved conductivities are obtained from the symmetries of the $N=5$ conductivity tensor given in Eq.~(\ref{eq:Sigma5}).

\section{Even-layer systems with higher $N$}
\label{app:higherN}

In this Appendix, we present explicit results for the hexalayer ($N=6$) and octalayer ($N=8$) systems, illustrating the general behavior of even-layer alternating-twist multilayers discussed in the main text. The susceptibilities are obtained by following the same procedure used for the tetralayer system. In addition, we propose a general relation between the outermost interlayer susceptibility and the total orbital susceptibility of even-layer systems in the flat-band regime, which is consistent with all explicit results presented in this work.

\subsection{Hexalayer ($N=6$) susceptibility}

For the first magic angle, $\theta^6_{1,m}=\beta^6_1\theta_m\approx1.89^{\circ}$, we obtain
\begin{align}
\chi_1/\chi_{\rm TBG}&=0.0036=\chi_5/\chi_{\rm TBG},\\
\chi_2/\chi_{\rm TBG}&=-0.0080=\chi_4/\chi_{\rm TBG},\\
\chi_3/\chi_{\rm TBG}&=0.0100,\\
\chi/\chi_{\rm TBG}&=0.0011 .
\end{align}
For the second magic angle, corresponding to $\theta^6_{2,m}=\beta^6_2\theta_m\approx1.31^{\circ}$, the
susceptibilities read
\begin{align}
\chi_1/\chi_{\rm TBG}&=0.0785=\chi_5/\chi_{\rm TBG},\\
\chi_2/\chi_{\rm TBG}&=-0.0435=\chi_4/\chi_{\rm TBG},\\
\chi_3/\chi_{\rm TBG}&=-0.0194,\\
\chi/\chi_{\rm TBG}&=0.0505 .
\end{align}
Finally, for the smallest magic angle, $\theta^6_{3,m}=\beta^6_3\theta_m\approx0.47^{\circ}$, we find
\begin{align}
\chi_1/\chi_{\rm TBG}&=1.4894=\chi_5/\chi_{\rm TBG},\\
\chi_2/\chi_{\rm TBG}&=1.1944=\chi_4/\chi_{\rm TBG},\\
\chi_3/\chi_{\rm TBG}&=2.1523,\\
\chi/\chi_{\rm TBG}&=7.5199 .
\end{align}
As in the case of the tetralayer, the effective magic angle with the highest $k$ yields only positive interlayer susceptibilities and has a larger total response compared to $\chi_{\mathrm{TBG}}$.

\subsection{Octalayer ($N=8$) susceptibility}

For the first magic angle, corresponding to $\theta^8_{1,m}=\beta^8_1\theta_m\approx1.97^{\circ}$, we obtain
\begin{align}
\chi_1/\chi_{\rm TBG}&=0.0008=\chi_7/\chi_{\rm TBG},\\
\chi_2/\chi_{\rm TBG}&=-0.0019=\chi_6/\chi_{\rm TBG},\\
\chi_3/\chi_{\rm TBG}&=0.0030=\chi_5/\chi_{\rm TBG},\\
\chi_4/\chi_{\rm TBG}&=-0.0034,\\
\chi/\chi_{\rm TBG}&=0.0002 .
\end{align}
For the second magic angle, given by $\theta^8_{2,m}=\beta^8_2\theta_m\approx1.61^{\circ}$, the
susceptibilities read
\begin{align}
\chi_1/\chi_{\rm TBG}&=0.0144=\chi_7/\chi_{\rm TBG},\\
\chi_2/\chi_{\rm TBG}&=-0.0194=\chi_6/\chi_{\rm TBG},\\
\chi_3/\chi_{\rm TBG}&=0.0067=\chi_5/\chi_{\rm TBG},\\
\chi_4/\chi_{\rm TBG}&=0.0027,\\
\chi/\chi_{\rm TBG}&=0.0061 .
\end{align}
The third effective magic angle corresponds to $\lambda^8_3=(\beta^8_3)^2=1$, and therefore
coincides with the TBG magic angle, $\theta_m$, giving
\begin{align}
\chi_1/\chi_{\rm TBG}&=0.1111=\chi_7/\chi_{\rm TBG},\\
\chi_2/\chi_{\rm TBG}&=0=\chi_6/\chi_{\rm TBG},\\
\chi_3/\chi_{\rm TBG}&=0=\chi_5/\chi_{\rm TBG},\\
\chi_4/\chi_{\rm TBG}&=-0.1111,\\
\chi/\chi_{\rm TBG}&=0.1111 .
\end{align}
This case is particularly interesting because the susceptibility is finite only in the outermost and central interlayer spaces, while it vanishes completely in the remaining ones.

Finally, for the smallest magic angle, associated with
$\theta^8_{4,m}=\beta^8_4\theta_m\approx0.35^{\circ}$, we obtain
\begin{align}
\chi_1/\chi_{\rm TBG}&=1.5404=\chi_7/\chi_{\rm TBG},\\
\chi_2/\chi_{\rm TBG}&=1.3546=\chi_6/\chi_{\rm TBG},\\
\chi_3/\chi_{\rm TBG}&=2.5459=\chi_5/\chi_{\rm TBG},\\
\chi_4/\chi_{\rm TBG}&=1.8896,\\
\chi/\chi_{\rm TBG}&=12.771 .
\end{align}

The explicit results presented above for the hexalayer and octalayer systems satisfy the general relation
\begin{align}
\chi_{\{1,N-1\}}^N(k)
=\lambda_k^N\chi^N(k)
=\left(\beta_k^N\right)^2\chi^N(k),
\end{align}
when the corresponding effective TBG subsystem is in the flat-band regime. Based on all cases analyzed in this work, we conjecture that this relation holds for arbitrary even-layer alternating-twist multilayers in the flat-band regime. The Mathematica code used to generate the analytical results presented in this work is publicly available on Zenodo as the \textit{Alternating-Twist Multilayer Graphene Symbolic Toolkit} (https://doi.org/10.5281/zenodo.21455129).

\newpage
\bibliography{biblio}

\end{document}